\documentstyle[aps,preprint,eqsecnum,epsfig]{revtex}
\begin{document}
\draft 
\tighten
\date{\today}

\title{Axion Radiation from Strings}

\author{C. Hagmann$^{1}$, S. Chang$^{2}$ and P. Sikivie$^{3}$}
\address{
$^{1}$Lawrence Livermore National Laboratory, Livermore, CA 94550}
\address{
$^{2}$Department of Physics, Purdue University, W. Lafayette, IN 47907}
\address{
$^{3}$Department of Physics, University of Florida, Gainesville, FL 32611}

\maketitle

\begin{abstract}
This paper revisits the problem of the string decay contribution to the
axion cosmological energy density.  We show that this contribution is 
proportional to the average relative increase when axion strings decay of 
a certain quantity $N_{\rm ax}$ which we define.  We carry out numerical 
simulations of the evolution and decay of circular and non-circular string 
loops, of bent strings with ends held fixed, and of vortex-antivortex pairs 
in two dimensions.  In the case of string loops and of vortex-antivortex 
pairs, $N_{\rm ax}$ decreases by approximately 20\%.  In the case of bent
strings, $N_{\rm ax}$ remains constant or increases slightly.  Our results 
imply that the string decay contribution to the axion energy density is of
the same order of magnitude as the well-understood contribution from vacuum 
realignment.  
\end{abstract}

\pagebreak

\narrowtext

\section{Introduction}

The axion was proposed over two decades ago as a solution to the 'Strong 
CP Problem', i.e. to explain why QCD conserves the discrete symmetries P 
and CP in spite of the fact that the Standard Model as a whole violates
those symmetries. It is the quasi-Nambu-Goldstone boson \cite{wewi78} 
associated with the spontaneous breakdown of a U$_{\rm PQ}(1)$ symmetry 
which Peccei and Quinn postulated \cite{pecc77}. The properties of the 
axion depend mainly on one unknown parameter, the magnitude of the vacuum 
expectation value $v_a$ which breaks U$_{\rm PQ}(1)$.  The mass of the 
axion and its couplings are inversely proportional to $v_a$. The 
(zero-temperature) mass is given by:
\begin{equation}
m_a  \simeq 6~\mu{\rm eV}~\cdot N \cdot {10^{12} {\rm GeV} \over v_a}~~\ .
\label{axmass}
\end{equation}
$N$ is a strictly positive integer that describes the color anomaly of 
U$_{\rm PQ}(1)$.  The combination of parameters $f_a \equiv {v_a \over N}$
is usually called the "axion decay constant".

A priori $v_a$, and therefore $m_a$, is a free parameter of the theory.
However, it is severely constrained by accelerator experiments and
astrophysical arguments.  Combined, the accelerator and astrophysical 
constraints imply $m_a\lesssim 10^{-2}{\rm eV}$ \cite{review}.  In addition, 
there is a {\it lower} limit on the axion mass from the requirement that 
axions do not overclose the Universe.  The axion cosmological energy density 
receives contributions from vacuum realignment \cite{abbo83}, from string 
decay \cite{davi85,hara87,davi89,hagm91,bash94,hagm99,yama00}, and from wall
decay \cite{hagm91,lyth92,naga94,chan99}.  All three contributions increase 
with decreasing axion mass.  The contribution whose size has been most 
controversial, and which is the topic of this paper, is that from string 
decay.  

Axion models contain strings because they have a spontaneously broken 
U(1) symmetry, namely U$_{\rm PQ}(1)$.  The latter is a global symmetry 
and hence axion strings are {\it global} strings.  The strings are only
present when the axion is massless, but such is indeed the case 
in the very early universe, when the temperature is much larger than 1 GeV.  
For the purposes of this paper we may describe the dynamics of axions and 
axion strings by the action density:
\begin{equation}
{\cal L} = \frac{1}{2} \partial_{\mu} \phi^\dagger \partial^{\mu} \phi
-\frac{\lambda}{4}(\phi^\dagger\phi-v_a^2)^2~~\ ,
\label{lagr}
\end{equation}
where $\phi \equiv \phi_1 + i \phi_2$ is a complex scalar field.  
${\cal L}$ is invariant under the U$_{\rm PQ}(1)$ symmetry: 
$\phi \to e^{i \alpha} \phi$.  The symmetry is spontaneously broken 
by the vacuum expectation value
\begin{equation}
\phi = v_a~e^{i{a \over v_a}}
\label{vev}
\end{equation}
where $a$ is the axion field.  A static axion string stretched along 
the $z$ axis is the field configuration:
\begin{equation}
\phi = v_a~f({r \over \delta})~e^{i \theta}~~~\ ,
\label{str}
\end{equation}
where $(z,r,\theta)$ are cylindrical coordinates,   
$\delta \equiv {1 \over \sqrt{\lambda} v_a}$ is the string core size, and 
$f({r \over \delta})$ is a function which goes to zero when $r \to 0$, 
approaches one for $r \gg \delta$, and solves equations of motion
which follow from Eq. (\ref{lagr}).  The energy per unit length of the 
string is
\begin{eqnarray}
\tau &=& \int d^2x ~[{1 \over 2} \vert \vec{\nabla} \phi \vert^2 +
{\lambda \over 4} (\phi^\dagger \phi - v_a^2)^2] \nonumber\\ 
&\simeq& \int_{\rho > \delta} d^2x ~{v_a^2 \over 2}~ 
\vert \vec{\nabla} e^{i\theta} \vert ^2 = \pi v_a^2 \ln({L \over \delta})
\label{tension}
\end{eqnarray}
where $L$ is an infra-red cutoff. The R.H.S. of Eq.(\ref{tension}) neglects 
the contribution to $\tau$ from the string core.  The string core contribution 
is of order $\pi v_a^2$, i.e. it is smaller than the contribution (1.5) of 
the field outside the string core by the factor $\ln(L/\delta)$.  In the 
situations of interest to us, $\ln({L \over \delta}) \simeq 67$.  Indeed, 
for axion strings in the early universe, the infra-red cutoff is provided by 
the presence of neighboring strings.  $L$ is then of order the distance 
between strings which, as we will see, is of order the horizon scale.  The 
relevant time is the start of the QCD phase transition, and hence
$L \sim  10^{-7}$ sec.  On the other hand, 
$\delta^{-1} \sim v_a \sim 10^{12}$ GeV.

Eq. (\ref{tension}) is valid for many but not all axion models.  In
general
\begin{equation}
\tau \simeq \pi ({v_a \over K})^2 \ln({L \over \delta})
\label{tengen}
\end{equation}
where $K$ is a model-dependent integer defined as follows.  Let $\alpha$ be 
the angle conjugate to the PQ charge. $K$ is the factor by which the period 
in $\alpha$ of the manifold of vacuum expectation values of the model is 
reduced by the presence of exact continuous symmetry transformations, such 
as gauge transformations.  In the PQWW model \cite{pecc77,wewi78} $K=2$, 
whereas in the DFSZ \cite{dine81} and KSVZ \cite{kims79} models $K=1$.

Axion strings appear as topological defects in the early universe when 
U$_{\rm PQ}(1)$ becomes spontaneously broken by the vacuum expectation 
value (\ref{vev}), at a temperature of order $v_a$.  The phase transition 
where this happens is called the PQ phase transition.  We assume in this 
paper that no inflation occurs after the PQ phase transition.  If there 
is inflation after the PQ transition, the axion field gets homogenized 
over enormous distances and the axion strings are 'blown away'.  In that 
case there is no string, nor wall, decay contribution to the axion 
cosmological energy density.  

At first the strings are stuck in the plasma \cite{hara87}, but soon the 
plasma becomes sufficiently dilute that the strings move freely and acquire 
relativistic speeds.  The strings are present till the axion mass turns on 
at the QCD phase transition.  The critical time is $t_1$ defined by
$m_a(T(t_1)) t_1 = 1$, where $m_a(T)$ is the temperature-dependent axion
mass. One finds \cite{abbo83}
\begin{equation}
t_1 \simeq 2 \cdot 10^{-7} {\rm sec} 
\left({f_a \over 10^{12} {\rm GeV}}\right)^{1 \over 3}~~\ .
\label{t1}
\end{equation}
The corresponding temperature is:
\begin{equation}
T_1 \simeq 1 {\rm GeV} 
\left({10^{12} {\rm GeV} \over f_a}\right)^{1 \over 6}~~\ .
\label{T1}
\end{equation}
All strings become the edges of $N_d$ domain walls at $t_1$, where $N_d$
is the number of degenerate vacua of the axion model \cite{siki82}.  It 
is related to $N$ by 
\begin{equation}
N_d = {N \over K}
\label{Nd}
\end{equation}
where $K$ is the same factor as appears in Eq.(\ref{tengen}). If 
$N_d \geq 2$, the universe becomes domain wall dominated.  The 
resulting cosmology is inconsistent with observation.  If $N_d = 1$, 
the walls bounded by string are unstable and decay into axions 
\cite{vile82}.  We recently gave a detailed discussion of the 
wall decay contribution to the axion cosmological energy density 
\cite{chan99}.

In this paper, our goal is to determine the present energy density 
$\rho_a^{\rm str}(t_0)$ in axions that were radiated by axions strings 
between the PQ phase transition and time $t_1$.  In section 2, we analyse 
the problem theoretically.  We express $\rho_a^{\rm str}(t_0)$ in terms 
of quantities $\xi, \chi$ and $\bar {r}$ which parametrize the main 
sources of uncertainty.  Most of the past debate has focused on 
$\bar {r}$ which is a functional of the energy spectrum of axions 
emitted by strings.  The remaining sections of the paper report on 
our estimates of $\bar {r}$ using numerical simulations.  Section 3 
describes our simulations of the motion and decay of circular and 
non-circular string loops.  Section 4 does the same for bent strings.  
Section 5 describes our simulations of the motion and annihilation of 
vortex-antivortex pairs in 2 (space) dimensions.  The behaviour of the
vortex-antivortex system can be predicted by analytical methods in the 
regime where the vortex and antivortex do not overlap.  Thus the 
vortex-antivortex pair provides an interesting case study where 
theory and simulation can be confronted with each other.  In section 6 
we summarize our results and compare them to the previous simulations by 
two of us \cite{hagm91}, to those of Battye and Shellard \cite{bash94}, 
and those of Yamaguchi, Kawasaki and Yokoyama \cite{yama00}.

\section{Theoretical analysis}

As we shall see, the axions radiated by strings become non-relativistic
soon after $t_1$.  Since each axion contributes $m_a$ to the energy today,  
our focus is on determining their {\it number} density $n_a^{\rm str}(t)$.  

Axions are radiated by collapsing string loops and by oscillating wiggles 
on long strings.  By definition, long strings stretch across the horizon. 
They move at relativistic speeds and intersect one another.  When strings 
intersect, there is a high probability of reconnection, i.e. of rerouting 
of the topological flux \cite{shel87}.  Because of such 'intercommuting', 
long strings produce loops which then collapse freely.  In view of this
efficient decay mechanism, the average density of long strings is 
expected to be of order the minimum consistent with causality, namely 
one long string per horizon.  Hence the energy density in long strings:
\begin{equation}
\rho_{\rm str}(t) = \xi {\tau \over t^2}  \simeq 
\xi \pi ({v_a \over K})^2 {1 \over t^2} \ln ({t \over \delta})~~\ ,
\label{strden}
\end{equation}
where $\xi$ is a parameter of order one. 

The equations governing the number density $n_a^{\rm str}(t)$ of axions 
radiated by axion strings are \cite{hara87}
\begin{equation}
{d\rho_{\rm str}\over dt} = -2 H\rho_{\rm str} - 
{d\rho_{{\rm str}\rightarrow a}\over dt}
\label{11}
\end{equation}
and
\begin{equation}
{dn_a^{\rm str}\over dt} = - 3H n_a^{\rm str} + {1\over \omega (t)}
{d\rho_{{\rm str}\rightarrow a}\over dt}
\label{12}
\end{equation}
where $\omega (t)$ is defined by:
\begin{equation}
{1\over \omega (t)} = {1\over {d\rho_{{\rm str}\rightarrow a}\over dt}} \int 
{dk \over k} {d\rho_{{\rm str}\rightarrow a}\over dt~dk}\ .
\label{13}
\end{equation}
$k$ is wavevector magnitude.  ${d\rho_{{\rm str}\rightarrow a}\over dt}(t)$
is the rate at which energy density gets converted from strings to axions 
at time $t$, and ${d\rho_{{\rm str}\rightarrow a}\over dt~dk}(t,k)$ is the 
spectrum of the axions produced.  $\omega(t)$ is therefore the average 
energy of axions radiated in string decay processes at time $t$.  The 
term $-2 H \rho_{\rm str} = + H \rho_{\rm str} - 3 H \rho_{\rm str}$ in 
Eq. (\ref{11}) takes account of the fact that the Hubble expansion both 
stretches $(+H\rho_{\rm str})$ and dilutes $(-3H\rho_{\rm str})$ long
strings.  Integrating Eqs. (\ref{strden} - \ref{12}), setting 
$H = {1 \over 2t}$, and neglecting terms of order one versus terms of 
order $\ln ({t \over \delta})$, one obtains
\begin{equation}
n_a^{\rm str}(t) \simeq 
{\xi \pi v_a^2 \over K^2 t^{3 \over 2}} \int_{t_{\rm PQ}}^t 
dt^\prime~{\ln({t^\prime \over \delta}) \over 
t^{\prime {3 \over 2}} \omega(t^\prime)}~~\ ,
\label{14}
\end{equation}
where $t_{\rm PQ}$ is the time of the PQ transition. 

The central question is: what is the average energy $\omega(t)$ of axions 
radiated in string decay processes at time $t$?  Axions are radiated by 
wiggles on long strings and by collapsing string loops.  Consider a 
process which starts at $t_{\rm in}$ and ends at $t_{\rm fin}$, and which 
converts an amount of energy $E$ from string to axions.  $t_{\rm in}$ and 
$t_{\rm fin}$ are both of order $t$.  The number of axions radiated is 
\begin{equation}
N = \int dk {dE \over dk}(t_{\rm fin}) {1 \over k}
\label{Na}
\end{equation}
where ${dE \over dk}(t_{\rm fin})$ is the energy spectrum of the $\phi$ 
field in the final state.  The average energy of axions emitted is:
\begin{equation}
\omega = {E \over N}~~\ .
\label{om}
\end{equation}
It is useful to define the quantity \cite{hagm91}
\begin{equation}
N_{\rm ax}(t) \equiv \int dk {dE \over dk}(t) {1 \over k}~~\ .
\label{Nax}
\end{equation}
The final value of $N_{\rm ax}$ is $N$.  The initial value of 
$N_{\rm ax}$ is determined in terms of $E$ by the fact that the 
energy stored in string has spectrum 
${dE \over dk} (t_{\rm in}) \sim {1 \over k}$.  This spectral 
shape - equal energy per unit logarithmic wavevector interval - 
is implied by Eq. (1.5).   If $\ell \equiv {E \over \tau}$ is the 
length of string converted to axions, we have 
\begin{equation}
{dE \over dk}(t_{\rm in}) = \pi ({v_a \over K})^2 \ell {1 \over k}
\label{Estr}
\end{equation}
for $k_{\rm min} < k < k_{\rm max}$ where $k_{\rm max}$ is of order 
${2 \pi \over \delta}$ and $k_{\rm min}$ of order ${2 \pi \over t}$.  
$t$ plays the role of $L$ in Eq. (1.5).  Hence
\begin{equation}
N_{\rm ax}(t_{\rm in}) = \pi ({v_a \over K})^2  \ell
\int_{k_{\rm min}}^{\delta^{-1}} dk {1 \over k^2} = 
{E \over \ln({t \over \delta}) k_{\rm min}}
\label{Naxin}
\end{equation}
Combining Eqs. (\ref{Na} - \ref{Naxin}) yields
\begin{equation}
{1 \over \omega} = {r \over \ln({t \over \delta}) k_{\rm min}}
\label{ome}
\end{equation}
where $r$ is the relative change in $N_{\rm ax}(t)$ during the process 
in question:
\begin{equation}
r \equiv {N_{\rm ax}(t_{\rm fin}) \over N_{\rm ax}(t_{\rm in})}~~\ .
\label{r}
\end{equation}
$k_{\rm min}$ is of order ${2\pi \over L}$ where $L$ is the loop 
size in the case of collapsing loops, and the wiggle wavelength in 
the case of bent strings. $L$ is at most of order $t$ but may be 
substantially smaller than that if the string network has a lot of 
small scale structure.  To parametrize our ignorance in this matter, 
we define $\chi$ such that the suitably averaged 
$k_{\rm min} = \chi {2 \pi \over t}$. Combining Eqs. (\ref{14}) and 
(\ref{ome}) we find:
\begin{equation}
n_a^{\rm str}(t) \simeq {\xi \bar{r} \over \chi} {v_a^2 \over K^2 t}~~\ ,
\label{na}
\end{equation}
where $\bar{r}$ is the weighted average of $r$ over the various processes
that convert string to axions.

Let us show that the set of all axions that were radiated between 
$t_{\rm PQ}$ and $t$ have spectrum ${d n_a \over dk} \sim {1 \over k^2}$ 
for ${1 \over t} \lesssim k \lesssim {1 \over \sqrt{t t_{\rm PQ}}}$,
irrespective of the shape of ${d\rho_{{\rm str}\rightarrow a}\over
dt~dk}$.  Indeed scaling implies 
\begin{equation}
{d\rho_{{\rm str}\rightarrow a}\over dt~dk} (t,k) =
{d\rho_{{\rm str}\rightarrow a}\over dt} (t) f(t k) t \ ,
\label{spec}
\end{equation}
where the unknown function $f(u)$ is normalized such that
\begin{equation}
\int_0^\infty f(u) du = 1 \ .
\label{norm}
\end{equation}
We will only assume that $f(u)$ has appreciable support near $u=1$.
Eqs. (\ref{strden}, \ref{11}) imply
\begin{equation}
{d\rho_{{\rm str}\rightarrow a}\over dt} (t) \simeq {1 \over t}
\rho_{\rm str} (t)~~\ ,
\label{dstrden}
\end{equation}
where "$\simeq$" indicates, as before, that terms of order one are 
neglected versus terms of order $\ln({t \over \delta})$.  Since 
axions free-stream after they are emitted, and $R \sim \sqrt{t}$
at the relevant epoch, we have:
\begin{equation}
{dn_a \over dk} (t,k) = \int_{t_{\rm PQ}}^t dt^\prime 
~{d\rho_{{\rm str}\rightarrow a}\over dt~dk} 
(k^\prime, t^\prime)~{1 \over k^\prime}~({t^\prime \over t})
\label{yep}
\end{equation}
with $k^\prime = k ({t \over t^\prime})^{1 \over 2}$.  The factor
${t^\prime \over t}$ accounts for the redshift and the volume expansion 
between $t^\prime$ and $t$.  Using Eqs. (\ref{spec}, \ref{dstrden}) and 
(\ref{strden}), one obtains
\begin{equation}
{dn_a \over dk} (t,k) \simeq {\xi \pi ({v_a \over K})^2 \over k^2 t^2}~
\int_{k \sqrt{t t_{\rm PQ}}}^{kt} du  f(u) \ln({u^2 \over k^2 t \delta})~~\ .
\label{des}
\end{equation}
For ${1 \over t} \lesssim k \lesssim {1 \over \sqrt{t t_{\rm PQ}}}$, the
integral in Eq. (\ref{des}) is only a slowly varying function of $k$ 
compared to the factor ${1 \over k^2}$.  That is the promised result.  

Since their momenta are of order $t_1^{-1}$ at time $t_1$, the axions 
become non-relativistic soon after they acquire mass.  Therefore, 
the string decay contribution to the axion energy density today is
\begin{equation}
\rho_a^{\rm str}(t_0) = m_a n_a^{\rm str}(t_1)({R_1 \over R_0})^3 \simeq
m_a {\xi \bar{r} \over \chi} {v_a^2 \over K^2 t_1} ({R_1 \over R_0})^3
\label{strcon}
\end{equation}
where ${R_1 \over R_0}$ is the ratio of scale factors between $t_1$ and 
today.  For comparison, the contribution from vacuum realignment is 
\cite{abbo83}
\begin{equation}
\rho_a^{\rm vac}(t_0) \simeq m_a {f_a^2 \over t_1} ({R_1 \over R_0})^3~~\ .
\label{vaccon}
\end{equation}
In terms of the critical energy density $\rho_c = {3 H_0^2 \over 8 \pi G}$,
the vacuum realignment contribution is:
\begin{equation}
\Omega_a^{\rm vac} \equiv {\rho_a^{\rm vac}(t_0) \over \rho_c} 
\simeq {1 \over 3} ({f_a \over 10^{12}{\rm GeV}})^{7 \over 6} 
({0.7 \over h})^2~~\ .
\label{Ome}
\end{equation}
As usual, $h$ parametrizes the present Hubble rate 
$H_0 = h \cdot 100 {{\rm km} \over {\rm sec} \cdot {\rm Mpc}}$.  
The contribution from wall decay is \cite{chan99}
\begin{equation}
\rho_a^{\rm d.w.}(t_0) \simeq m_a {6 \over \gamma} {f_a^2 \over t_1} 
({R_1 \over R_0})^3
\label{wallcon}
\end{equation}
where $\gamma$ is the average Lorentz factor of axions produced in the 
decay of walls bounded by string.  In simulations, we found $\gamma \sim 7$ 
for ${v_a \over m_a} \simeq 500$, but that $\gamma$ increases approximately 
linearly with $\ln({v_a \over m_a})$.  Extrapolation of this behaviour to 
the parameter range of interest, $\ln({v_a \over m_a}) \simeq 60$, yields 
$\gamma \sim 60$. It suggests that the wall decay contribution is subdominant 
relative to the vacuum realignment contribution.  

The ratio of the string decay and vacuum realignment contributions is:
\begin{equation}
{\rho_a^{\rm str}(t_0) \over \rho_a^{\rm vac}(t_0)} \simeq
{\xi \bar{r} N_d^2 \over \chi}~~\ .
\label{ratio}
\end{equation}
where we used Eq. (\ref{Nd}).  Each of the factors on the R.H.S.
deserves discussion:

$N_d ~~$  Almost surely one needs $N_d=1$ to avoid the cosmological 
disaster of an axion domain wall dominated universe.   It may be worth 
pointing out, however, that the domain wall problem can be avoided, 
in $N_d > 1$ models, by introducing an interaction which slightly lowers 
one of the $N_d$ vacua with respect to the others \cite{siki82}.  The 
lowest vacuum takes over after some time and the walls disappear before 
they dominate the energy density.  There is little room in parameter 
space for this to happen, but it is logically possible.  It is discussed 
in detail in ref. \cite{chan99}.

$\xi~~~$  In previous work \cite{hagm91,hagm99}, we set $\xi = 1$ on the 
argument that the number density of long strings should be close to the 
minimum consistent with causality, i.e. of order one long string per horizon.  
Battye and Shellard \cite{bash94} set $\xi \simeq 13$ because this describes 
the density of strings in simulations of {\it local} string networks in an 
expanding universe.  However, axion strings are {\it global} strings.  
Unlike global strings, local strings have all their energy located in the 
string core.  Also, they cannot dissipate their energy by emitting 
Nambu-Goldstone radiation as global strings do.  For these reasons, it is 
not obvious that global strings are as dense in the early universe as local 
strings would be.  In fact, M. Yamaguchi, M. Kawasaki and J. Yokoyama 
\cite{yama00} have done simulations of global string networks in an expanding 
universe and find $\xi \simeq 1$.

$\chi~~~$ $\chi$ and $\xi$ are related since the average interstring distance 
controls both.  On dimensional grounds, $\chi \sim \sqrt{\xi}$.  So, the 
effect of small scale structure in the axion string network partially cancels 
out in the RHS of Eq. (\ref{strcon}). In previous work \cite{hagm91,hagm99}, 
we have assumed $\chi \simeq 1$ but this could be off by a factor two or
so.

$\bar{r}~~~$  This is the unknown on which most of the past debate has 
focused.  Two basic scenarios have been put forth, which we call A and B.  
The question is: what is the spectrum of axions radiated by strings?  The 
main source is closed loops of size $L \sim t$.  Scenario A postulates 
that a bent string or closed loop oscillates many times, with period of 
order $L$, before it has released its excess energy and that the 
spectrum of radiated axions is concentrated near  ${2 \pi \over L}$.  In
that case one has $\bar{r} \sim \ln({t_1 \over \delta}) \simeq 67$.  
Scenario B postulates that the bent string or closed loop releases its 
excess energy very quickly and that the spectrum of radiated  axions is 
${dE \over dk} \sim {1 \over k}$ with a high frequency cutoff of order 
${2 \pi \over \delta}$ and a low frequency cutoff of order 
${2 \pi \over L}$.  In scenario B, the initial and final spectra 
${dE \over dk}$ of the energy stored in the axion field are qualitatively 
the same and hence $\bar{r} \sim 1$.  In scenario A, the string decay
contribution dominates over the vacuum realignment contribution by the
factor $\ln({t_1 \over \delta})$, whereas, in scenario B, the
contributions from string decay and vacuum realignment have the same 
order of magnitude. 

Computer simulations offer a way to try and estimate $\bar{r}$.  It should 
be kept in mind however that present day technology limits lattice sizes 
to approximately $256^3$ in 3 dimensions (3D), and $4096^2$ in 2D.  Hence 
the ratio of loop/core size that can be investigated is limited to 
$\ln({L \over \delta}) \simeq$ 3.5 in 3D, and 5 in 2D, whereas 
$\ln({L \over \delta}) \simeq 67$ in the situations of physical interest. 
It is therefore important to investigate the dependence, if any, of the
results of computer simulations on $\ln({L \over \delta})$ over the small
range that can be investigated.  The following three sections report on
our simulations of circular and non-circular string loops, bent strings,
and vortex-antivortex pairs.  

\section{String Loop Simulations}

We simulated the motion and decay of circular loops initially at rest, and 
of non-circular loops with angular momentum.  The initial configurations 
are set up on large ($\sim 10^7$ points) Cartesian grids, and time-evolved 
using the finite-difference equations derived from Eq. (\ref{lagr}).  FFT 
spectrum analysis of the kinetic and gradient energies during the collapse 
yields $N_{\rm ax}(t)$, and hence $r$.

\subsection{Circular Loops}

Because of azimuthal symmetry, circular loops can be studied in $\rho-z$ 
space.  The $z$-axis is perpendicular to the plane of the loop.  By mirror 
symmetry, the problem can be further reduced to one quarter-plane.  The 
static axion field outside the string core is \cite{hagm91}
\begin{equation}
a(\rho,z) =  \frac{v_a}{2} \Omega(\rho,z)
\label{solan}
\end{equation}
in the infinite volume limit, where $\Omega$ is the solid angle subtended 
by the loop.  We use as initial configuration the outcome of a relaxation 
routine starting with Eq.(\ref{solan}) outside the core and
\begin{equation}
\phi(r) = \tanh(0.58 {r \over \delta}) e^{i\theta}
\end{equation}
within the core.  Here, $r$ is the distance to the string center, and 
$\theta$ is the polar angle about the string.  The configuration inside 
the core is held fixed during the relaxation.  The relaxation and  
subsequent dynamical evolution are done with reflective Neumann-type 
boundary conditions. A step size $dt$ = 0.2 was used for the time 
evolution.  The total energy was conserved to better than 1\%. 

We call $R_0$ the initial loop radius.  Fig. 1 shows $R(t)$ for collapsing 
loops.  While they collapse, the loops reach speeds close to the speed 
of light.  Figs. 2a through 2e show successive snapshots of the string
core as it speeds toward the origin.  It is increasingly Lorentz 
contracted.  A lattice effect is observed when the Lorentz contracted core 
size becomes comparable to the lattice spacing.  This effect consists of a
``scraping'' of the string core on the underlying grid, with dissipation
of the kinetic energy of the string into high frequency axion radiation.  
We chose $\lambda$ small enough to avoid this phenomenon.  For the large 
($R_0 = 2400$) 2D circular loop simulations, $\lambda \lesssim 0.01$ is 
required.

In most cases, the loops collapse without rebound.  However, for the range 
of parameters $80 \lesssim R_0/\delta \lesssim190$, a bounce occurs.  It 
is shown in Fig. 1 (solid line) for the case $R_0 = 2400$ and $\delta = 15.8$. 
Figs. 2f through 2l show snapshots of the axion field configuration during 
the bounce.  Note that the orientation of the string switches during the 
bounce, i.e. a left-oriented loop bounces into a right-oriented one, or 
vice-versa.

Spectrum analysis of the fields was performed by expanding the 
gradient plus kinetic energy
\begin{equation}
E_{\rm kin+grad} = \int_{-L_Z}^{L_z} dz \int_0^{L_\rho} 2\pi \rho d\rho~ 
({1 \over 2} \dot{\phi}^\dagger \dot{\phi} + 
{1 \over 2} \vec{\nabla} \phi^\dagger \cdot \vec{\nabla} \phi)
\end{equation}
using
\begin{eqnarray}
\dot{\phi}=  \sum_{mn}a_{mn} J_0(k_m\rho)\,\cos(k_n(z+L_z)) \nonumber\\
\nabla_z\phi=\sum_{mn}b_{mn} J_0(k_m\rho)\,\sin(k_n(z+L_z)) \nonumber\\
\nabla_\rho\phi=\sum_{mn}c_{mn} J_0(k_m\rho)\,\cos(k_n(z+L_z))
\end{eqnarray}
where the $k_m$ and $k_n$ satisfy $J_0(k_m L_\rho)=0$ and 
${\rm sin}(2 k_n L_z)=0$ respectively. The wavevector magnitude of mode 
$(m,n)$ is $k_{mn} \equiv \sqrt{k_m^2+k_n^2}$. Fig. 3 shows the power 
spectrum at $t=0$ and, after the collapse, at $t=3000$.  At both times, 
it exhibits an almost flat plateau, consistent with $dE/dk\propto 1/k$.  
The high frequency cutoff of the spectrum is increased however at the 
later time, as might be expected since the core gets Lorentz contracted 
during the collapse.  The evolution of 
$N_{\rm ax}= {\displaystyle \sum_{mn}{E_{mn} \over k_{mn}}}$ was studied 
for various values of $R_0/\delta$.  As shown in Fig. 4, we observe a 
marked decrease of $N_{\rm ax}$ during the collapse, of order $20 \%$ 
independent of $R_0/\delta$ in the investigated range 
$3.6 \leq \ln(R_0/\delta) \leq 5.0$.  In the previous circular
loop simulations by two of us \cite{hagm91}, in which $\ln (R_0/\delta)$ 
ranges from 2.5 to 3.4, it was also found that $r \simeq 0.8$.  

\subsection{Non-circular loops}

Circular loops are a special case and their behaviour may be untypical 
of loops in general.  To address this concern, we performed simulations 
of non-circular loops as well.  The initial condition of the loop is given 
by its initial position $\vec{r}(s,t_0)$ and velocity $\vec{v}(s,t_0)$, 
where $s$ parametrizes distance along the loop.  Only the transverse 
part of $\vec{v}(s,t_0)$ has physical meaning.  We determine the initial 
axion field using
\begin{equation}
a(\vec{x},t_0) = {v_a \over 2} \Omega(\vec{x})\{\vec{r}(s,t_0)\}
\label{solang}
\end{equation}
where $\Omega(\vec{x})\{\vec{r}(s,t_0)\}$ is the solid angle subtended 
by the loop as seen from $\vec{x}$.  The axion field $a(\vec{x},t_0+dt)$ 
a short time $dt$ later is similarly determined by calculating the 
solid angle subtended by the loop located at 
$\vec{r}(s,t_0) + dt~\vec{v}(s,t_0)$.  This yields the initial time 
derivative of the axion field $\dot{a}(\vec{x},t_0)$.  

Scenario A postulates that an axion string behaves like a Nambu-Goto
(NG) string to lowest order in a perturbative expansion in powers of 
${1 \over \ln({L \over \delta})}$ where $L$ is the string size.  The 
most general solution to the NG equations of motion is 
\cite{ggrt73}:
\begin{equation}
\vec{r}(s,t) = {R_0 \over 2} [\vec{a}(\sigma_-) + \vec{b}(\sigma_+)]
\label{solut}
\end{equation}
where $\sigma_\pm \equiv (s \pm t)/R_0$, $s$ is proportional to the 
energy in the string between the points labeled $s$ and $s=0$, and 
$\vec{a}(\sigma)$ and $\vec{b}(\sigma)$ are arbitrary functions of 
period $2\pi$, which satisfy 
$\vec{a}^\prime(\sigma)^2 =\vec{b}^\prime(\sigma)^2 = 1$.  $L = 2\pi R_0$ 
is the total proper length of the string loop, i.e. 
$\vec{r}(s+L,t) = \vec{r}(s,t)$.  It can be shown \cite{kibb82} that a 
NG string loop of length $L$ has a motion which is periodic in time with 
period $L/2$, and that an initially static NG string loop collapses to a 
doubled-up line after half a period.

A considerable literature \cite{kibb82,turo84,burd85,chen88} is devoted 
to the problem of finding non-intersecting NG loop solutions.  The authors
are mainly motivated by issues related to the cosmic string scenario of 
large scale structure formation.  However, self-intersection of string 
loops is also relevant to the question whether scenario A or B is 
more likely to be correct.  Intercommuting (self-intersection with 
reconnection) favors scenario B because it causes loop sizes to shrink, 
and hence the average energy of radiated axions to increase.  As discussed in 
Ref. \cite{hagm91}, if the probability $p$ of intercommuting per oscillation 
is larger than of order ${1 \over \ln({L \over \delta})} \simeq 1.4\%$, the 
spectrum of axions radiated by the original loop, its two daughters, four 
grand-daughters, and so on, will be qualitatively the same as in scenario B, 
independently of assumptions on the spectrum of radiation from any 
single loop. 

A particular two-parameter set of solutions \cite{kibb82,turo84} is given 
by:
\begin{eqnarray}
x(s,t)&=&\frac{R_0}{2}\left((1-\alpha)\,\sin \sigma_-
           +\frac{1}{3}\alpha\,\sin 3\sigma_-
           +\sin\sigma_+ \right) \nonumber\\
y(s,t)&=&\frac{R_0}{2}\left(-(1-\alpha)\,\cos \sigma_-
           -\frac{1}{3}\alpha\,\cos 3\sigma_-
           -\cos \psi\,\cos \sigma_+ \right) \nonumber\\
z(s,t)&=&\frac{R_0}{2}\left(-2\sqrt{\alpha(1-\alpha)}\,\cos \sigma_-
           -\sin \psi\,\cos \sigma_+ \right)~~\ . 
\label{KT}
\end{eqnarray}
with $\alpha\in (0,1)$ and $\psi\in (-\pi,\pi)$.  In a large region of
$(\alpha, \psi)$ parameter space, the loop does not self-intersect 
\cite{chen88}.  

We performed 21 simulations of non-circular loops using Eqs.(\ref{KT}) 
as initial conditions, on a $256^3$ lattice with periodic boundary 
conditions.  Each simulation took approximately 1 week to run.  A 
variety of $(\alpha,\psi)$ and $\lambda$ values were chosen.  Figure 5
shows the collapse of a non-circular loop projected onto the $xz$ plane.  
Here, and in all cases tried, the loop size shrinks to zero in one go, 
without oscillation or rebound.  Standard Fourier techniques were used 
for the spectrum analysis, and $N_{\rm ax}$ was computed as a function 
of time.  In all but two of the cases tried, $N_{\rm ax}$ decreases 
while the loop collapses.  The $r$ values depend upon $\alpha,~\psi$ 
and $\lambda$, and cover the range 0.6 to 1.07.  Figure 6 shows 
$N_{\rm ax}(t)$ for $R_0 = 72, \alpha= 0.7, \psi = \pi/2$ and $\lambda$ =
0.0125, 0.025, 0.05 and 0.1.  Some of the largest $r$ values (1.06, 0.94,
0.90 and 0.89 respectively) were obtained in these simulations.  In them, 
and in all cases where $\ln({R_0 \over \delta})$-dependence was tested, 
$r$ decreases with increasing $\ln({R_0 \over \delta})$.  It appears 
however that the second derivative is decreasing and that $r$ may reach
an asymptotic value for $\ln({R_0 \over \delta}) \sim 3$.  For 
$\alpha = 0.7,~\psi = \pi/2$, the asymptotic value would be of order 0.9. 

Nine simulations were done for $\lambda = 0.05$ and $R_0 = 72$
(hence $\ln({R_0 \over \delta})$ = 2.78) and a variety of
$(\alpha,~\psi)$ values.  The average of $r$ over this set is 0.77.  

\section{Bent String Simulations}

We have also carried out simulations of oscillating bent strings with 
ends held fixed.  The oscillation amplitude decreases as the string loses 
energy to axion radiation.  We choose the $z$ axis parallel to the direction 
of the string when straight.  $L_x L_y L_z$ is the size of the 
box.  Initial configurations describing static, sinusoidally shaped strings 
were prepared using the ansatz
\begin{equation}
\phi(x,y,z)=\tanh({.58\,r \over \delta})\,\exp\left(i\arctan\left(
\frac {y-y_0}{x-x_0-A_0\sin(\frac {2 \pi z}{\Lambda})}\right)\right) 
\end{equation}
where $r=\sqrt{(x-x_0-A_0\sin(\frac {2 \pi z}{\Lambda}))^2+(y-y_0)^2}$, 
$A_0$ is the initial amplitude, $(x_0,y_0) = ({L_x \over 2}, {L_y \over 2})$ 
is the equilibrium position of the string and $\Lambda = L_z$ is the string 
wavelength.  We used square boxes ($L_x = L_y$) in all cases.

The field outside the core was thoroughly relaxed before dynamical time 
evolution was begun.  Neumann boundary conditions were imposed on the four 
side faces $(x=0,L_x,$ and $y=0,L_y)$ and periodic boundary conditions on
the endfaces $(z=0,L_z)$.  The kinetic + gradient energy in the $\phi$
field was spectrum analyzed at regular time intervals during the dynamical 
evolution.  We took care to avoid finite volume and discrete space-time 
effects.  To minimize finite volume (i.e. boundary) effects, 
$L_x, L_y \geq 4 L_z$ is needed.  To avoid discretization effects, 
$\lambda \leq 0.4$ and time step $dt \leq 0.2$ are needed.  When the latter 
conditions are satisfied, no scraping of the string on the underlying 
lattice is observed and total energy is conserved to better than one part 
in $10^3$.

We performed runs for a variety of box sizes, initial amplitudes $A_0$ 
and core sizes $\delta = {1 \over \sqrt{\lambda}}$.  Fig. 7 shows the 
amplitude $A(t)$ as a function of time for initial amplitudes 
$A_0 = 20$ and 10, on a 256*256*64 lattice and $\lambda = 0.2$.  For 
$t \lesssim 250$, the damped oscillator behaviour of $A(t)$ is very 
smooth and regular.  For $t \gtrsim 250,~A(t)$ is less regular because 
the string is being driven by radiation which was emitted in the first 
couple of oscillations and which returned to the string's location after 
reflection by the sidewalls,

Fig. 7 shows that, in general, $A(t)$ is not proportional to $A_0$.  
After two oscillations the amplitudes are of the same order, 
$A(t = 140) \simeq 6$, even though the initial amplitudes differ by a
factor two.  We confirm the following rule, already stated in 
ref. \cite{hagm91}: oscillations of initial amplitude $A_0$ much larger
than $\Lambda /10$ decay rapidly, in one or two oscillations, till 
$A(t) \lesssim \Lambda/10$.  After that the string is more weakly damped.

Fig. 8 shows $N_{\rm ax}(t)$ for $A_0 = 30,~20,~10$ on a 256*256*64
lattice and $\lambda = 0.2$.  $N_{\rm ax}$ increases slightly, of 
order 1\%, and then oscillates about an average value.  It is not clear 
to us whether this slight increase of $N_{\rm ax}$ is a real effect
because we were unable to satisfy ourselves that it happens in the
infinite volume limit.  However, let us analyze it as a real 
effect.  To this end, we need a formula for $r$ in the case of bent
strings, when only part of the string decays into axions.  The fraction 
of string that has decayed between the initial time $t_{\rm in}$ and 
the final time $t_{\rm fin}$ is given by the fractional change 
$1-{\ell_{\rm fin} \over \ell_{\rm in}}$ in the length $\ell$ of 
the string between those times.  For a sinusoidally shaped string 
the length is determined in terms of the amplitude $A$ by
\begin{equation}
\ell = \int_0^{L_z} dz 
\sqrt{1 + ({2 \pi A \over \Lambda} \cos {2 \pi z \over \Lambda})^2}~~\ .
\label{length}
\end{equation}
Let us call $N_{\rm ax}^\prime$ the value of $N_{\rm ax}$ restricted to
that fraction of string which decays into axions between times $t_{\rm in}$ 
and $t_{\rm fin}$.  We have 
\begin{equation}
N_{\rm ax}^\prime(t_{\rm in}) = 
(1-{\ell_{\rm fin} \over \ell_{\rm in}}) N_{\rm ax}(t_{\rm in})
\label{Npin}
\end{equation}
and  
\begin{equation}
N_{\rm ax}^\prime(t_{\rm fin}) = N_{\rm ax}(t_{\rm fin}) - 
{\ell_{\rm fin} \over \ell_{\rm in}} N_{\rm ax}(t_{\rm in})~~\ .  
\label{Npfin}
\end{equation}
Hence, we estimate $r$ for bent string simulations using the formula
\begin{equation}
r = {N_{\rm ax}^\prime(t_{\rm fin}) \over N_{\rm ax}^\prime(t_{\rm in})}
= {N_{\rm ax}(t_{\rm fin}) - 
{\ell_{\rm fin} \over \ell_{\rm in}} N_{\rm ax}(t_{\rm in}) \over 
(1-{\ell_{\rm fin} \over \ell_{\rm in}}) N_{\rm ax}(t_{\rm in})}~~\ .
\label{bentr}
\end{equation}
Table 1 shows the $r$ values for $A_0 =$ 30, 20, 10 and 
$\lambda =$ 0.05, 0.1, 0.2 on a 256*256*64 lattice.  $r$ increases with 
decreasing $A_0$, reaching values of order 1.12 for $A_0 = 10$.  However, 
for such small initial amplitudes, only a very small part of the energy
stored in the string gets released.  The larger amplitude 
($A_0  \sim 0.5 \Lambda$) simulations are more relevant because they 
are typical of the cosmological setting and because proportionately
more energy gets released in them.  For large initial amplitude, $r$ 
is close to one and has a tendency to decrease with increasing 
$\ln({L \over \delta})$.  In contrast, scenario A predicts that $r$ 
is of order $\ln({L \over \delta})$.  

\section{Vortex-Antivortex Annihilation}

We also studied the annihilation of vortex-antivortex pairs in 2D.
The Lagrangian is the same as before, Eq. (1.2).  Note however that in 
2D the field $\phi$ and its expectation value $v_a$ have dimension 
of (length)$^{-1/2}$, and $\lambda$ has dimension of (length)$^{-1}$.  
The energy stored in a vortex at rest is 
$E \simeq \pi v_a^2 \ln(L/\delta)$, which is analogous to 
Eq.(\ref{tension}).  The core size is 
$\delta = {1 \over \sqrt{\lambda}v_a}$ as before.  For a vortex-antivortex 
pair, the distance $R$ between them is the infra-red cutoff $L$.  The total 
energy in the pair is therefore
\begin{equation}
E(R) \simeq 2 \pi v_a^2 \ln({R \over \delta})
\label{pairen}
\end{equation}
when the vortex and antivortex are at rest.  

A vortex-antivortex pair has zero topological charge.  It annihilates 
by emitting Nambu-Goldstone (NG) radiation.  We may think of this 
process in 3D as the annihilation of an infinitely long straight string 
with an infinitely long parallel antistring.  This is not directly 
relevant to the problem considered here since long parallel axion 
strings are unlikely in the early universe.  However, because it is 2D, 
the process can be accurately simulated.  Moreover the behaviour of this 
relatively simple system can be predicted on theoretical grounds.  We will
argue below that the energy spectrum of NG radiation emitted by a spinning
vortex anti-vortex pair has the qualitative shape 
${dE \over dk} \sim {1 \over k}$. 

Simulations of vortex-antivortex annihilation were carried out previously 
in Refs. \cite{shel87,hech90}.  However, as far as we know, the present 
work is the first to spectrum analyze the NG radiation emitted in the 
decay.  Ref. \cite{yama98} presents simulations of the formation and 
evolution of vortices in an expanding 2+1 dimensional universe.

What should one expect the radiation spectrum to be?  It is known that, 
in 2+1 dimensions, the axion field is dual to a gauge field which couples 
to the vortex as if it were a particle with electric charge 
$e = \sqrt{2 \pi} v_a$.  This is the restriction to 2+1 dimensions 
of the well-known duality \cite{kalb74,witt85,hech90} relating the axion 
field in 3+1 dimensions to an anti-symmetric two-index gauge field 
$A_{\mu\nu}(x)$ which couples to the world-sheet of the axion string.  
Hence, as long as the vortex and anti-vortex are at greater distance 
from one another than the sum of their core sizes $(R > 2\delta)$, they 
behave like a pair of oppositely charged particles.  When the cores of 
the vortex and anti-vortex start to overlap, this description is no longer 
valid.  However, in the generic situation where the pair has angular 
momentum, most of the energy has already been dissipated into radiation 
by then.

The force between the vortex and anti-vortex is attractive and has 
magnitude
\begin{equation}
F(R) = {2\pi v_a^2 \over R}~~~\ .
\end{equation}
It is a manifestation of the aforementioned duality that this force can be 
thought of either as the gradient of the potential energy (\ref{pairen}) or 
as the Coulomb force between two oppositely charged particles, of charges 
$\pm \sqrt{2\pi} v_a$.  The acceleration of both vortex and antivortex 
is therefore:
\begin{equation}
a \simeq {2 \over R \ln({R \over \delta})}
\end{equation}
in the not highly relativistic regime, and neglecting the radiation
reaction due to the emission of NG particles.  The typical angular 
frequency of the radiation is $\omega \sim a \sim {1 \over R}$.  Hence 
its spectrum has the generic form:
\begin{equation}
{dE \over d\omega} = {dE \over dR} {dR \over d\omega}
\sim 2\pi v_a^2 {1 \over R} {1 \over \omega^2} \sim 
2\pi v_a^2 {1 \over \omega}~~~\ ,
\end{equation}
i.e. it is flat on a logarithmic scale.  The low frequency radiation is 
emitted first and the high frequency last.

A $2048^2$ lattice was initialized with the ansatz   
\begin{equation}
\phi(x,y)=\phi_0(x-x_1,y-y_1)\,\phi_0^\ast(x-x_2,y-y_2)
\end{equation}
where $(x_1,y_1)$ and $(x_2,y_2)$ are the locations of the vortex and 
antivortex respectively, and $\phi_0(x,y)$ is the field of a single vortex 
at rest.  A relaxation routine was applied to this configuration with the 
cores held fixed.  Periodic boundary conditions were used during the
relaxation and the subsequent dynamical evolution.  The vortex and
antivortex were given an initial relative velocity $v_0$ perpendicular 
to their line of sight.  The range of parameters simulated was $0< v_{0}<
0.6$ for the initial velocity and $5 < \delta < 30$ for the core size.  
Fig. 9 shows snapshots of a decaying vortex-antivortex system for 
$\lambda=0.005$ and $v_0=0.5$. The initial and final ${dE \over d\ln k}$ 
energy spectra are shown in Fig. 10.  The initial spectrum, that of the 
initial vortex-antivortex pair, is flat like that of an axion string.  
The final spectrum is that of the NG radiation after the vortex and 
antivortex annihilated.   It is also flat qualitatively, in agreement 
with the theoretical argument given above.  In all cases, the final 
spectrum is found to be somewhat harder than the initial spectrum.  
Thus, $N_{\rm ax}$  decreases ($r<1$) for all parameters investigated.

\section{Conclusions}

We carried out numerical simulations of the decay of axion string into 
axions, for a variety of initial configurations.  Our main goal is to 
estimate the factor $\bar{r}$ which appears in the expression for the 
axion cosmological energy density, Eq. (\ref{na}).  $r$ is the relative 
increase of $N_{\rm ax}$ during a decay process.  $\bar{r}$ is the 
average of $r$ over the various processes that contribute.  We simulated 
circular loops, non-circular loops and bent strings.  

The circular loop simulations were done in 2D, exploiting the axial 
symmetry.  This allowed us to reach $\ln({R_0 \over \delta}) \simeq 5$. 
We found that circular loops collapse in one go, except in the range 
$80 < R_0/\delta < 190$ where they bounce once.  As far as we know, the 
bounce phenomenon was not observed in previous simulations nor was 
it anticipated in theoretical investigations.  We found $r \simeq 0.8$ 
for circular loops, whether or not they bounce.  We did not find any 
dependence of $r$ on $\ln({R_0 \over \delta})$ over the range, 
$3.6 < \ln({R_0 \over \delta}) < 5.0$, of the simulations.  Our 
results are consistent with the previous simulations of circular 
loops in 3D by two of us \cite{hagm91}, which showed $r \simeq 0.8$ 
over the range $2.6 < \ln({R_0 \over \delta}) < 3.2$ . 

Twenty-one non-circular loop simulations were carried out using as initial
conditions a family of Kibble-Turok configurations parametrized by $\alpha$ 
and $\psi$. In the case of Nambu-Goto strings, such initial conditions yield 
periodic non self-intersecting motion for most of the parameter space.  In 
the simulations, the loops collapse in one go, without oscillation or bounce.  
$N_{\rm ax}$ decreases in almost all cases.  $r$ depends on $\alpha$,
$\psi$ and $\ln({R_0 \over \delta})$, and ranged from 0.60 to 1.07.  
$r$ decreases with increasing $\ln({R_0 \over \delta})$ but appears to be 
reaching a limiting value for $\ln({R_0 \over \delta}) \sim 3$.  Nine 
simulations were done for $\ln({R_0 \over \delta}) = 2.78$.  The average
of $r$ over this set is 0.77. 

The bent string simulations were done on lattices of size 256*256*64 
and 256*256*128 with the string in the direction of the shortest 
dimension.  We found that oscillations of initial amplitude $A_0$ much 
larger than $\Lambda/10$, where $\Lambda$ is the wavelength of the wiggle, 
decay rapidly, in one or two oscillations, till $A(t) \lesssim \Lambda/10$.  
After that, the string is more weakly damped.  This is consistent with 
what was found in ref. \cite{hagm91}. We find that $N_{\rm ax}$ increases
slightly, by an amount of order 1\%.  The $r$ values are listed in 
Table 1 for a representative set of parameter values.

Our simulations are inconsistent with scenario A which predicts that $r$ 
is of order $\ln({L \over \delta})$.  We find that $r$ is of order one and 
also that $r$ does not increase with increasing $\ln({L \over \delta})$
over the range investigated, $3 < \ln({L \over \delta}) < 5$.  Our bent 
string simulations are also inconsistent with scenario A in that $A(t)$ 
is not proportional to the initial amplitude $A_0$.

Battye and Shellard \cite{bash94} have carried out bent string
simulations.  Their string configuration is similar to ours.  
They conclude that the string emits radiation mainly at twice the 
oscillation frequency of the string and that the spectrum falls off 
exponentially for large $k$.  We do not find evidence of this.  To obtain
the spectrum of radiated axions they subtract the field of a static
straight string when the string is going through the equilibrium position.  
The subtracted field is assumed to be that of the radiated axions and is
spectrum analyzed.  It is unclear to us that this subtraction procedure is 
valid because it neglects retardation and Lorentz contraction effects 
associated with the fact that the string is moving when it is going through 
its equilibrium position.  

Yamaguchi, Kawasaki and Yokoyama \cite{yama00} carried out simulations
of a network of global strings in an expanding universe.  They found 
$\xi \simeq 1$ for the parameter which describes the average density 
of strings [Eq. (\ref{strden})].  We use their result below.  They find 
that the spectrum of radiated axions is softer than the  
${dE \over dk} \sim {1 \over k}$ spectrum of the energy stored in
strings, whereas we find in most cases that it is somewhat harder.
It must be kept in mind however that, because the Yamaguchi et
al. simulations describe a network of many strings, their effective 
value of $\ln({L \over \delta})$ is small compared to the range of 
$\ln({L \over \delta})$ values explored by the simulations described 
here.

To estimate the axion cosmological energy density, we use 
$\bar{r} \simeq 0.8$ as implied by our circular and non-circular loop 
simulations.  The contribution of bent long strings is expected to be 
much less important than that from closed loops.  We use $\xi \simeq 1$ 
based on the simulations of global string networks in an expanding
universe by Yamaguchi et al. \cite{yama00}, and set $N_d = 1$ and 
$\chi = 2^{\pm 1}$ as discussed in the Introduction.  Hence our estimate 
for the string decay contribution:
\begin{equation}
\Omega_a^{\rm str} \simeq {\xi \bar{r} \over \chi} \Omega_a^{\rm vac}
\simeq 0.27~2^{\pm 1} 
\left({f_a \over 10^{12} {\rm GeV}}\right)^{7 \over 6} 
\left({0.7 \over h}\right)^2~~\ .
\label{strest}
\end{equation}
For the reasons mentioned in the Introduction, the wall decay contribution 
is probably subdominant compared to the vacuum realignement and string
decay contributions.  Hence our estimate for the total density 
in cold axions today:
\begin{equation}
\Omega_a \simeq \Omega_a^{\rm vac} + \Omega_a^{\rm str} = (0.4~{\rm to}~0.9) 
\left({f_a \over 10^{12} {\rm GeV}}\right)^{7 \over 6}
\left({0.7 \over h}\right)^2~~\ .
\label{totest}
\end{equation}
This is relevant to the dark matter searches presently in progress
\cite{hagm98,mats91}, insofar that it suggests the range of axion masses 
for which axions are the dark matter of the universe.  Let us remind the 
reader that the above estimate is for the case where there is no inflation
after the PQ phase transition.  It also assumes that the axion to entropy 
ratio is constant from time $t_1$ till the present.  Various ways in which 
this assumption may be violated are discussed in the papers of
ref. \cite{entrop}.  

Finally, we simulated the annihilation of vortex-antivortex pairs in 2D.
We find that the spectrum of radiated axions has the qualitative shape 
${dE \over dk} \sim {1 \over k}$.  This is consistent with our theoretical 
analysis of the system.  We find that $N_{\rm ax}$ decreases in this 
case too.
 
\acknowledgements

One of us (P.S.) thanks the Aspen Center for Physics for its hospitality 
while he was working on this project.  This research was supported in part 
by DOE grant DE-FG02-97ER41029 at the University of Florida and by DOE
grant W-7405-ENG-048 at Lawrence Livermore National Laboratory.

%%%%%%%%%%%%%%%%%%%%%%%%%%%%%%%%%%%%%%%%%%%%%%%%%%%%%%%%%%%%%%%%%%%%%%%%%%

\begin{table}
\caption{$r$ values for bent string simulations on a 256*256*64
lattice. We used $L = 64$ for estimating $\ln({L \over \delta})$.}
\begin{tabular}{cccc}
 $\lambda$ & $\ln({L \over \delta})$ & $A_0$ & $r$  \\
\tableline 
 0.05   &  2.7  &  30  &   1.07      \\
 0.05   &  2.7  &  20  &   1.08      \\
 0.05   &  2.7  &  10  &   1.12      \\
 0.1    &  3.0  &  30  &   1.04      \\
 0.1    &  3.0  &  20  &   1.06      \\
 0.1    &  3.0  &  10  &   1.12      \\
 0.2    &  3.4  &  30  &   1.02      \\
 0.2    &  3.4  &  20  &   1.05      \\
 0.2    &  3.4  &  10  &   1.12      \\

\end{tabular}
\label{tbl1}
\end{table}

%%%%%%%%%%%%%%%%%%%%%%%%%%%%%%%%%%%%%%%%%%%%%%%%%%%%%%%%%%%%%

\begin{figure}
\begin{center}
\vspace{3cm}
\includegraphics[width=14cm]{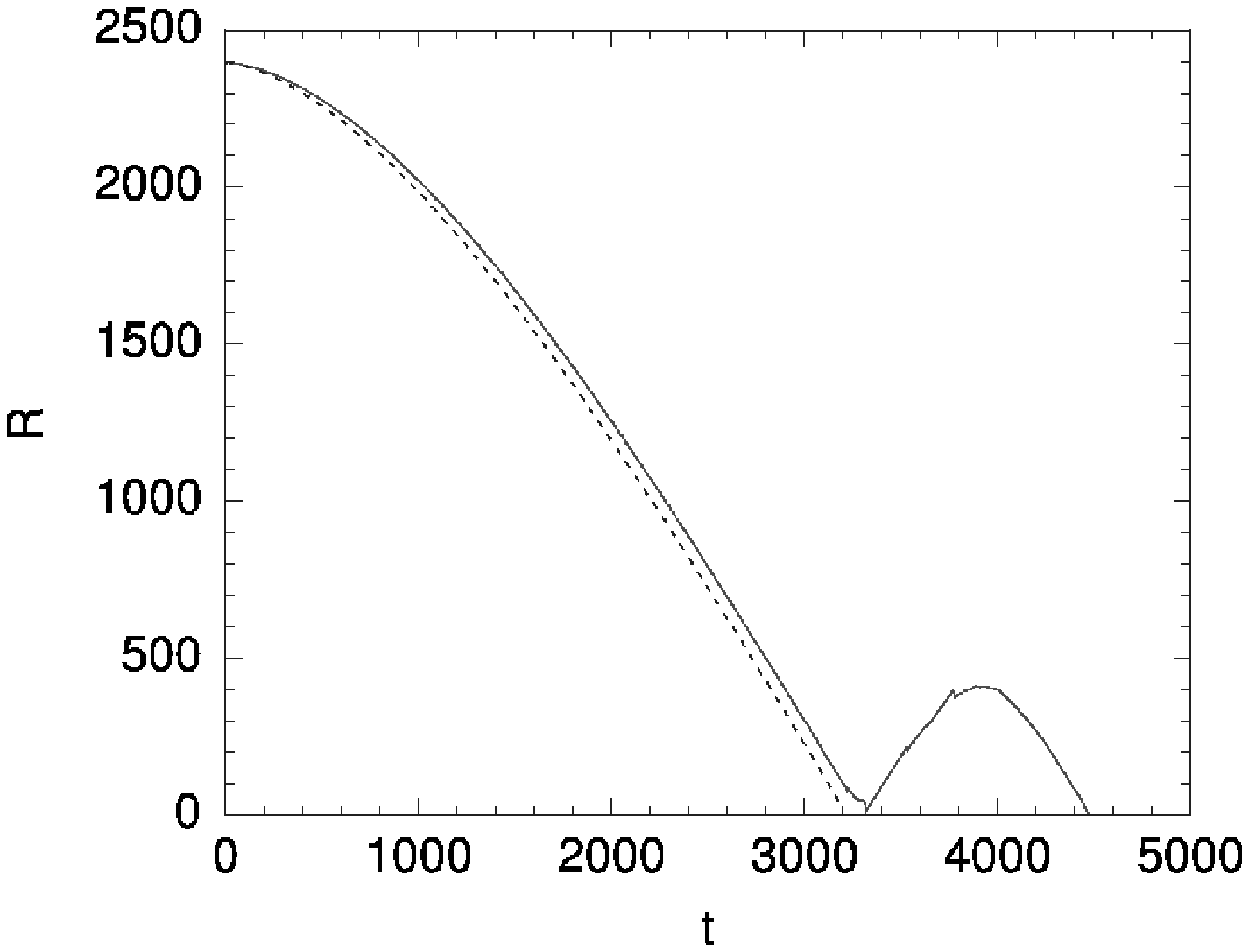}
\end{center}
\vspace{2cm}
\caption{Radius versus time of a collapsing circular loop, for 
$\lambda=0.001$ (dotted line) and $\lambda=0.004$ (solid line).}

\newpage
\begin{center}
\includegraphics[width=16cm]{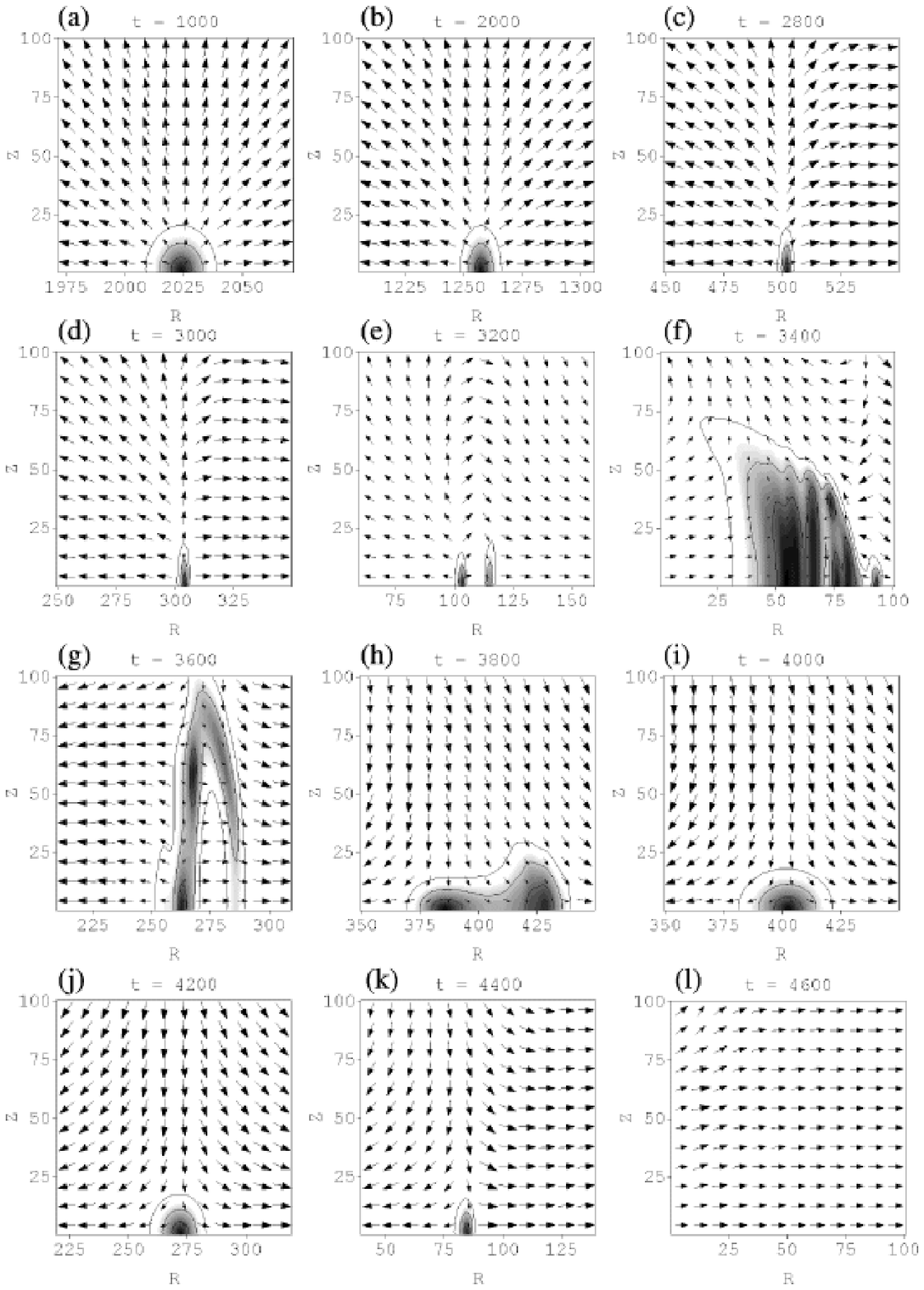}
\end{center}
\caption{Snapshots of a collapsing circular loop.  The arrows give the
direction of $(\phi_1,\phi_2)$ in internal space.  The grey-scale shows
the magnitude of $|\phi|$, white for $|\phi|=1$ and black for $|\phi|=0$.} 

\newpage
\begin{center}
\includegraphics[width=16cm]{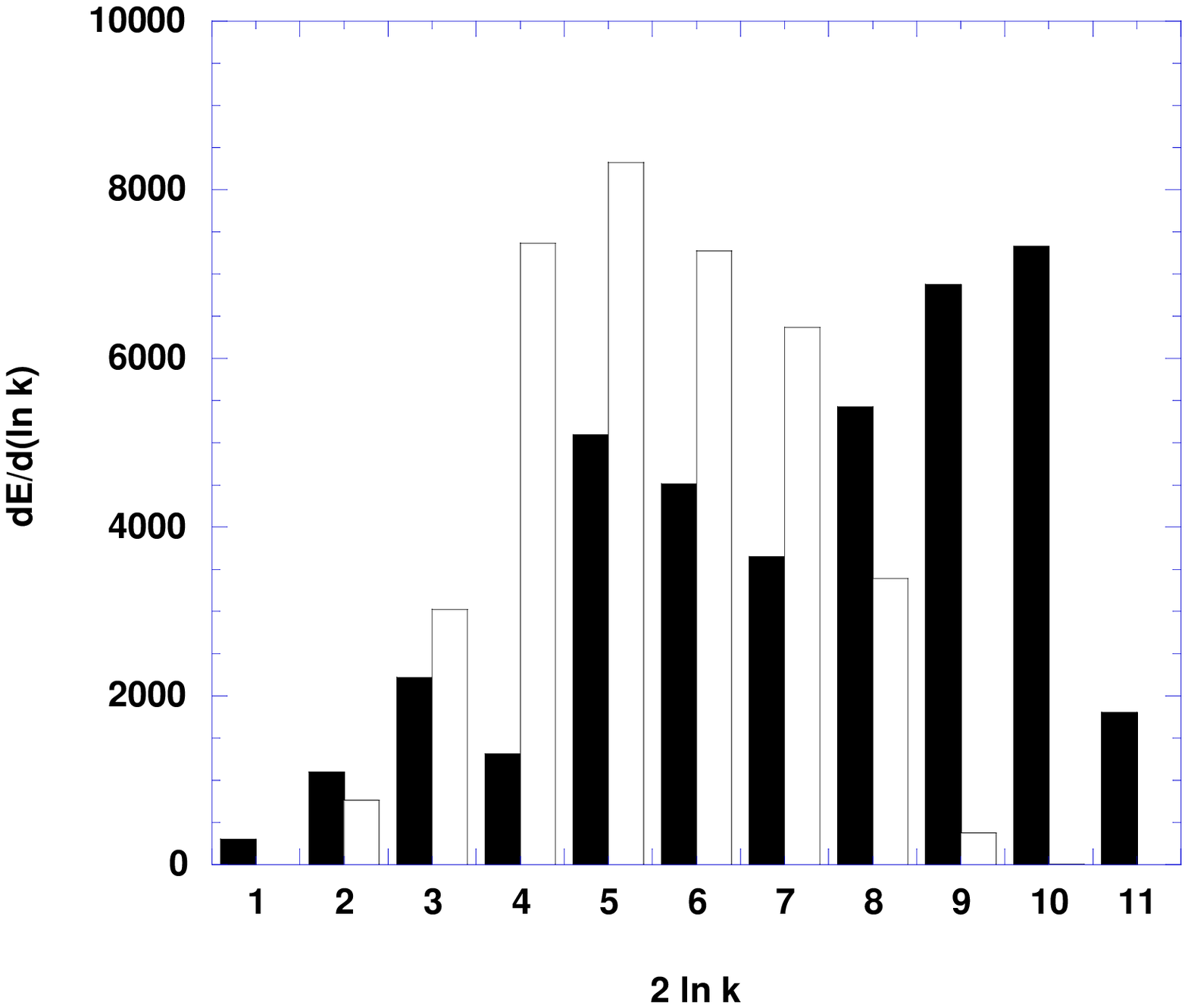}
\end{center}
\caption{Spectrum of decaying circular loop for $R_0=2400$, 
$\lambda=0.004$, at times $t=0$ (open boxes) and $t=5000$ (closed
boxes).}

\newpage
\begin{center}
\includegraphics[width=16cm]{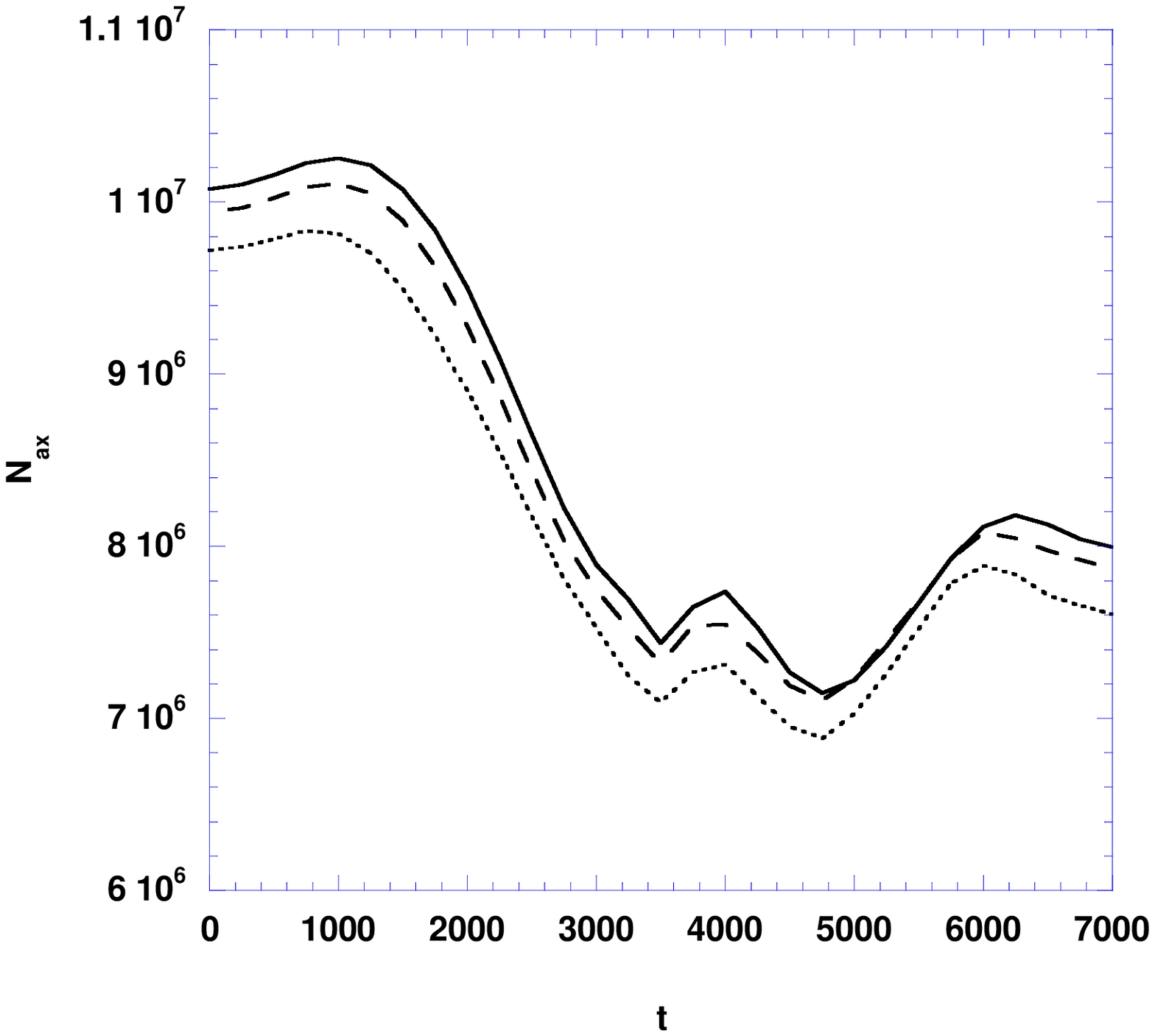}
\end{center}
\caption{$N_{\rm ax}$ versus time of a circular loop for $R_0=2400, 
~L_\rho=4000,~L_z=4096$ and $\lambda=0.004$ (solid), 0.001 (dashed), 
and 0.00025 (dotted).}

\newpage
\begin{center}
\includegraphics[width=16cm]{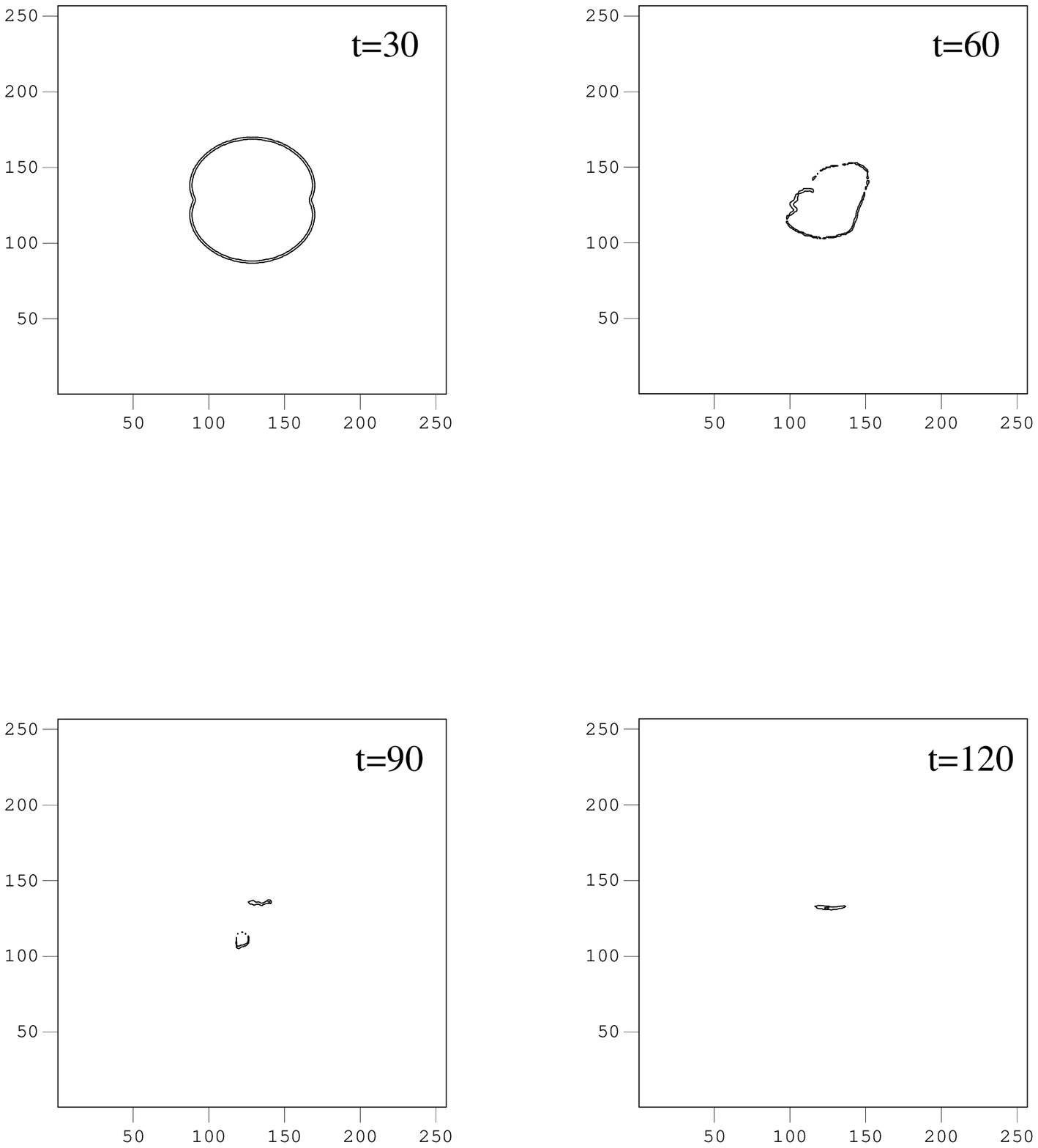}
\end{center}
\caption{Snapshots of a collapsing non-circular loop.  The simulation is 
carried out on a $256^3$ lattice with $\lambda = 0.05$. The initial 
conditions are given by Eq. (\ref{KT}) with $\alpha = 0.7, \psi = \pi/2, 
R_0 =72$.  The position of the string core is projeced onto the $xz$ plane.}

\newpage
\begin{center}
\includegraphics[width=16cm]{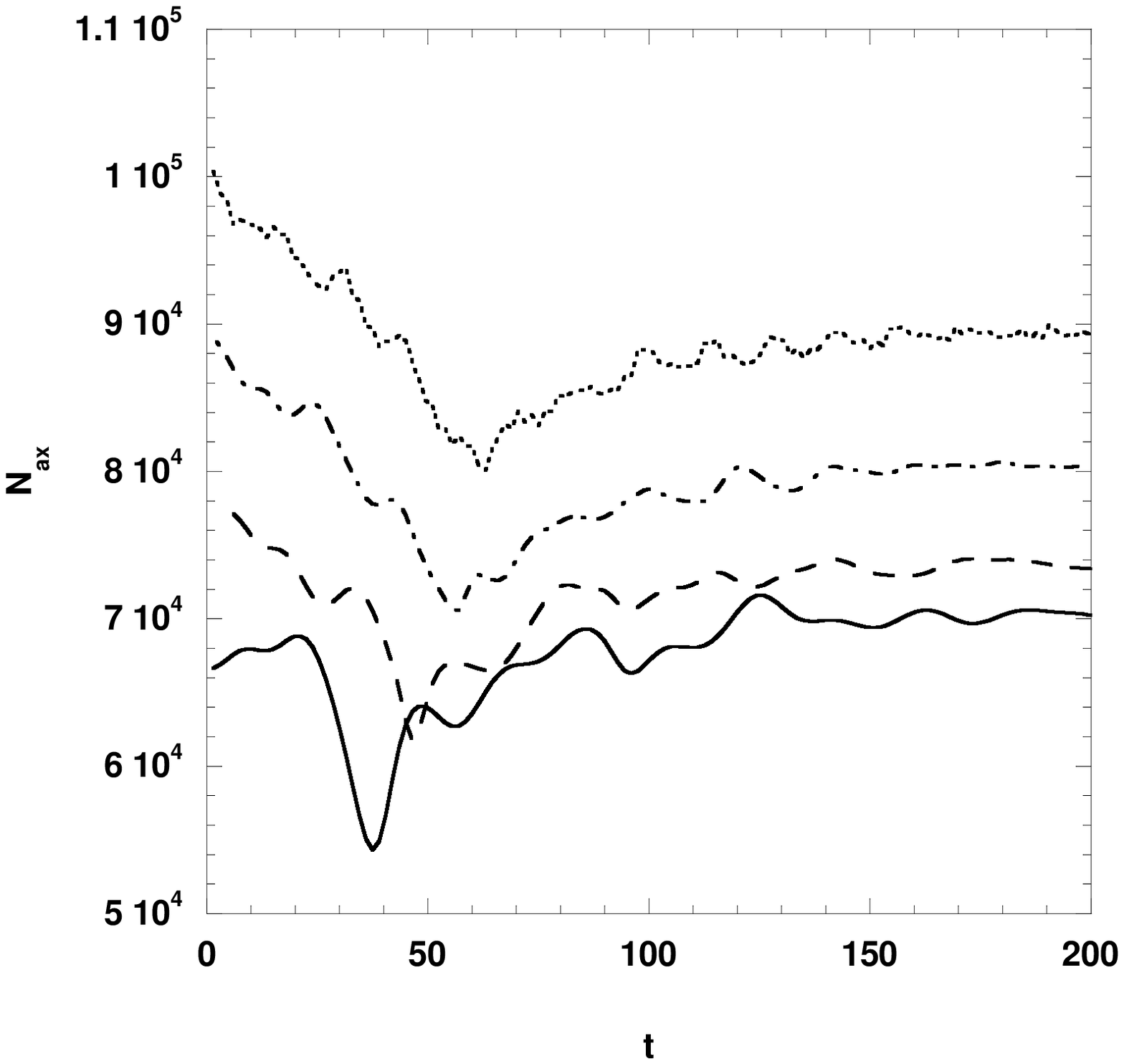}
\end{center}
\caption{$N_{\rm ax}$ versus time for four non-circular loop simulations 
on a $256^3$ lattice.  The initial conditions are given by Eq. (\ref{KT}) 
with $R_0 = 72, \alpha = 0.3, \psi = \pi/2$, and $\lambda = 0.1$ (dotted),
0.05 (dot-dashed), 0.025 (dashed) and 0.0125 (solid).}

\newpage
\begin{center}
\includegraphics[width=16cm]{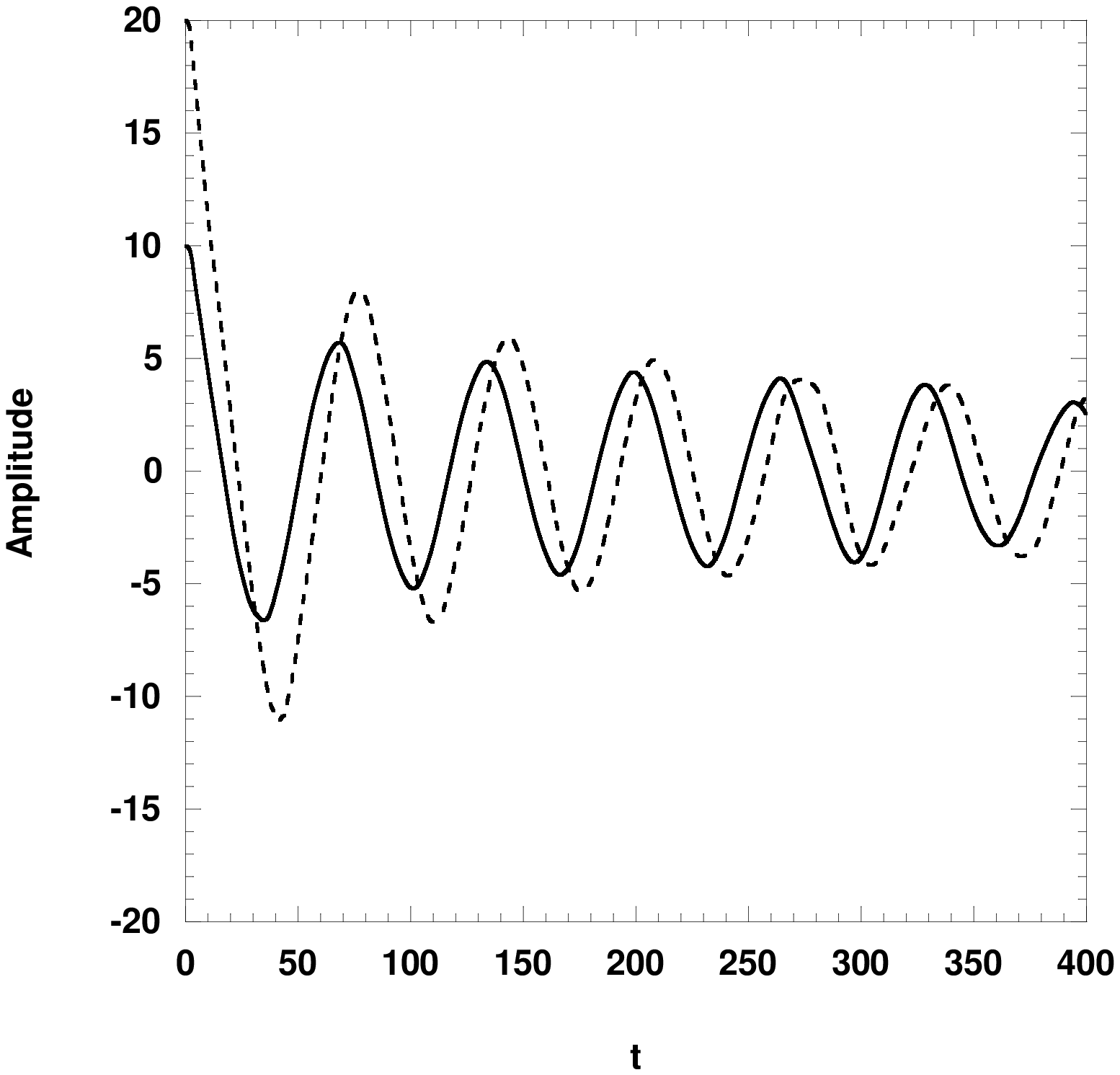}
\end{center}
\caption{String core position of oscillating bent strings versus time.  
The simulation is carried out on a $L_x L_y L_z = 256^2*64$ lattice 
with $\Lambda=64$, $\lambda=0.2$,and $A_0=20$ and 10.}

\newpage
\begin{center}
\includegraphics[width=16cm]{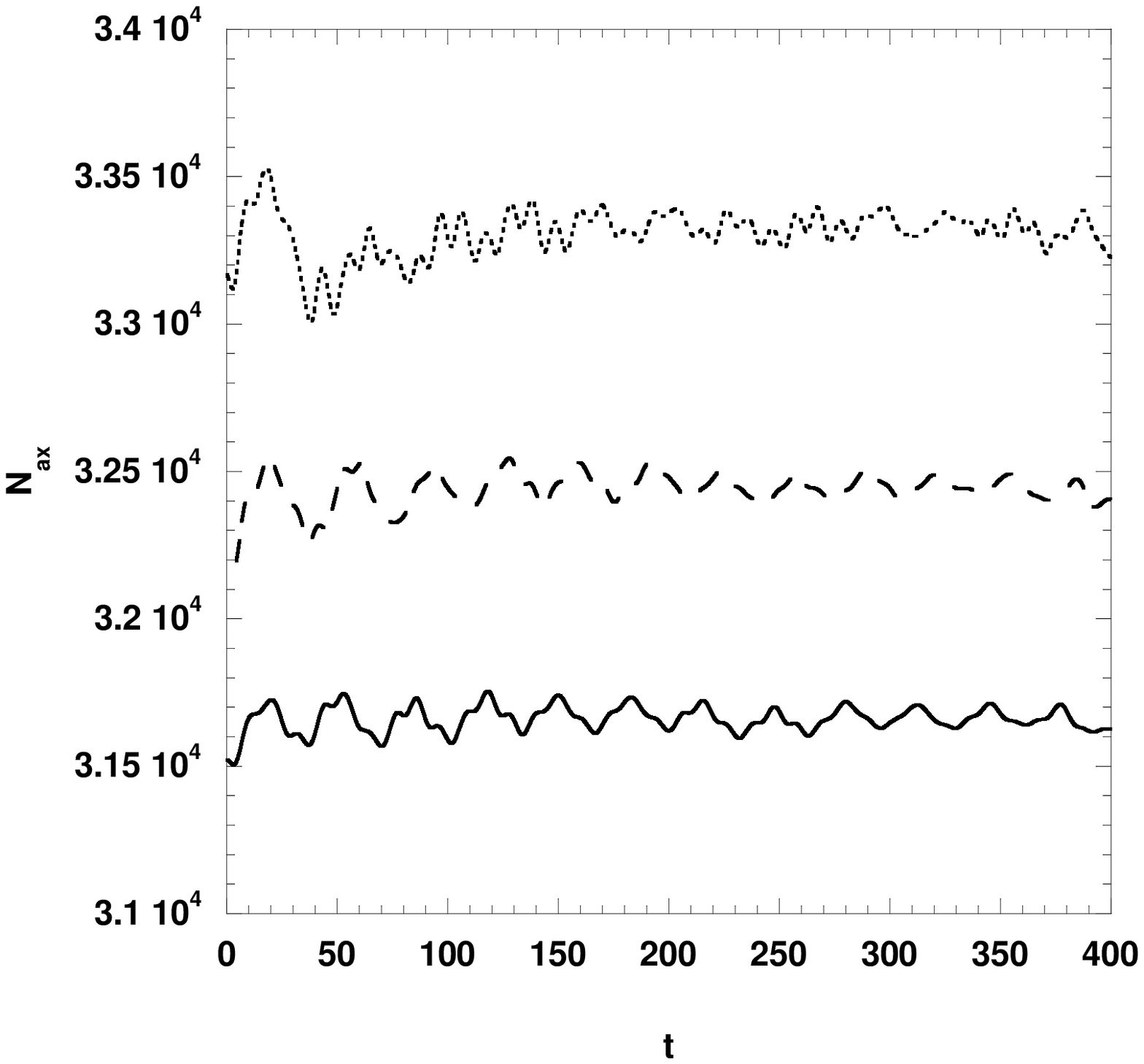}
\end{center}
\caption{$N_{\rm ax}$ for an oscillating bent string versus time. 
Here $L_x L_y L_z = 256^2*64$, $\Lambda=64$, $\lambda=0.2$ and
$A_0=30({\rm dotted}),~20({\rm dashed}),~10({\rm solid})$}.

\newpage
\begin{center}
\hspace{-2cm}\includegraphics[width=15cm]{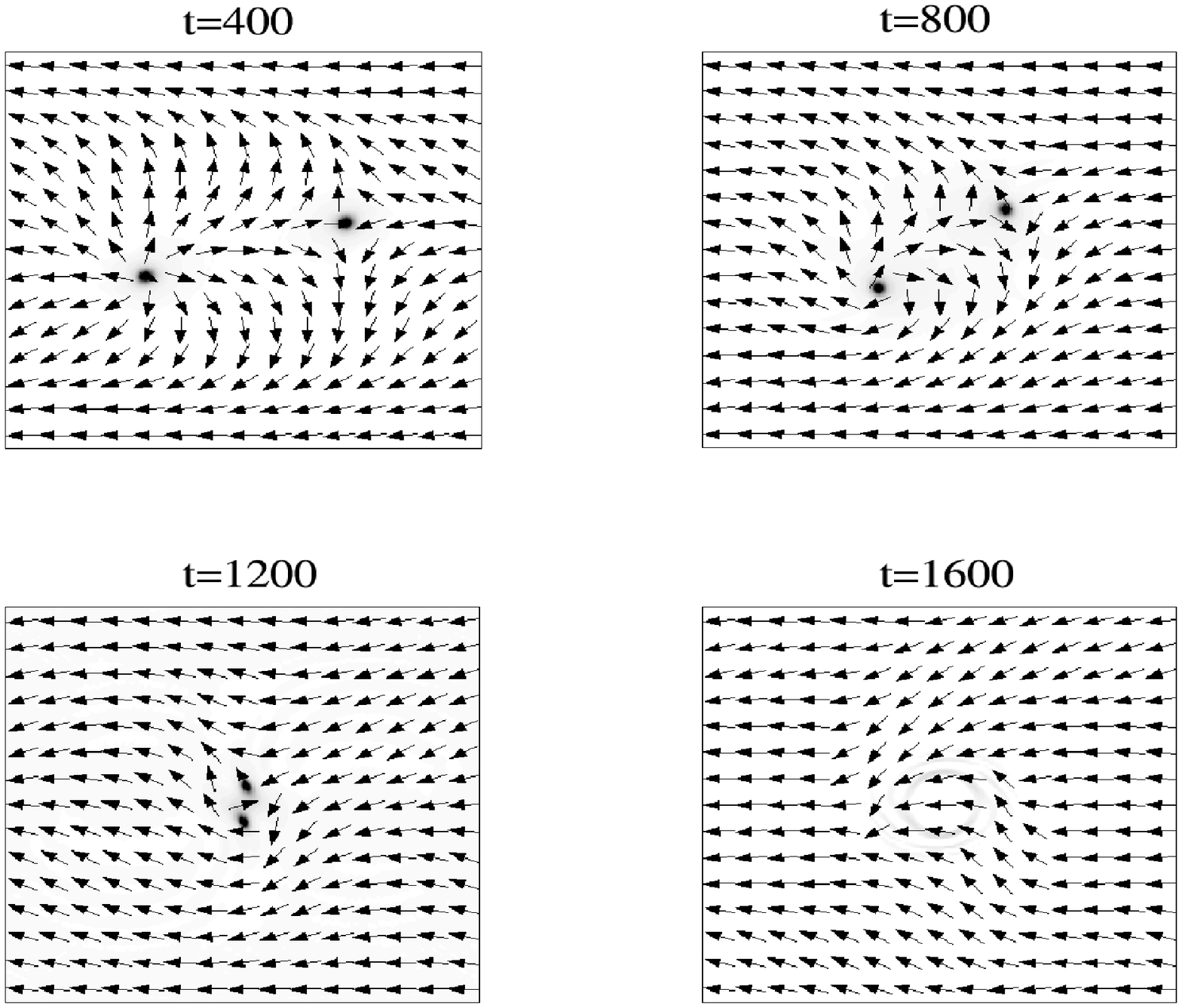}
\end{center}
\caption{Snapshots of collapsing vortex-antivortex pairs.  The 
simulation is carried out on a $2048^2$ lattice with $\lambda=0.005$ 
and $v_0=0.5$.  The initial separation between the vortex and 
anti-vortex is 1000 lattice units.}

\newpage
\begin{center}
\includegraphics[width=16cm]{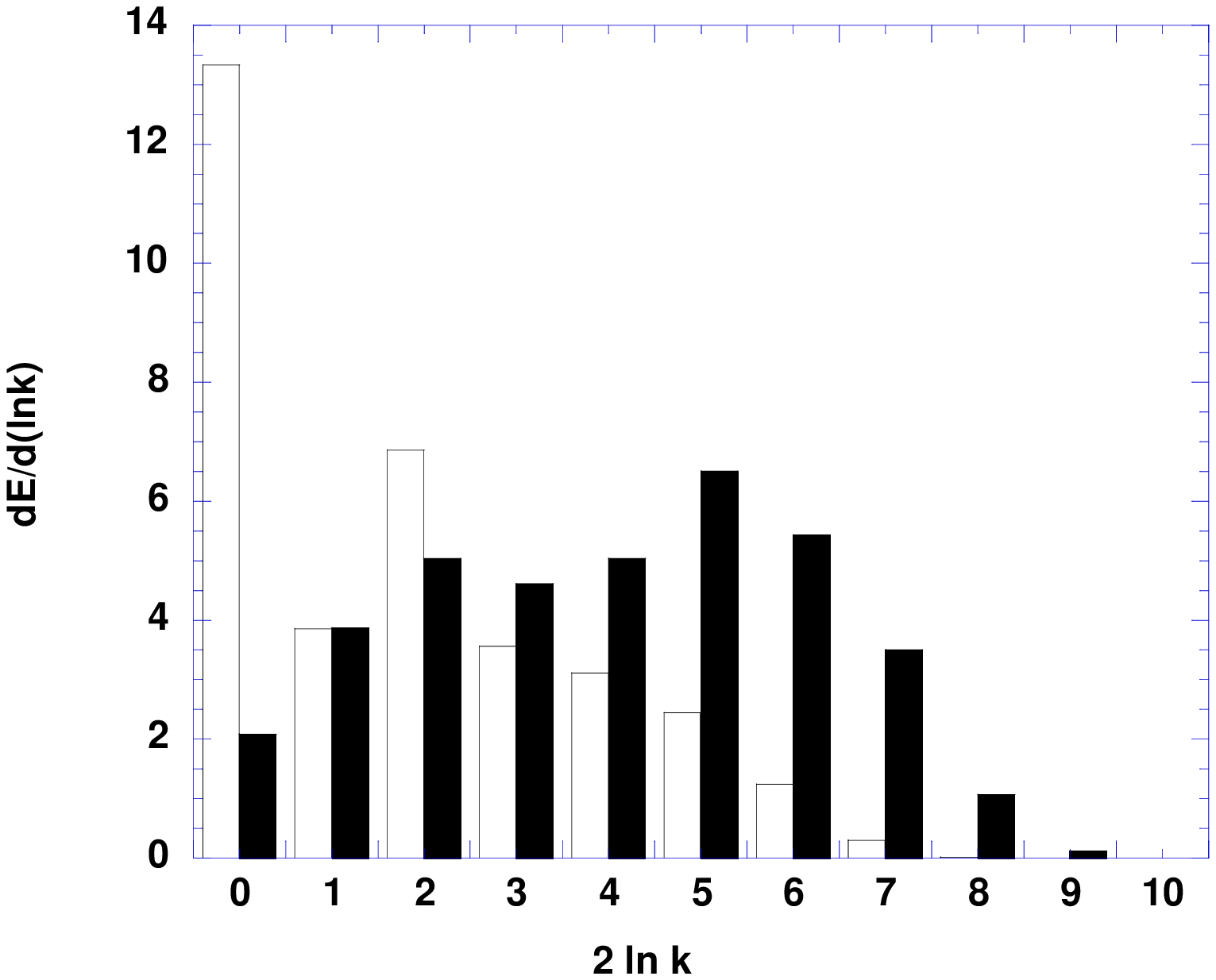}
\end{center}
\caption{Spectra of a vortex-antivortex pair at time times $t=0$ 
(open boxes) and $t=2000$ (closed boxes).  The parameter values are 
the same as in Fig. 9.}

\end{figure}

\end{document}